\documentclass[letterpaper]{aa501}

\usepackage{rotating}
\usepackage{graphics}
\usepackage{psfig}

\def\ea{et~al.}

\def \hi {H\,{\sc i~}}

\def\NH{$N_{\rm HI}$}
\def\kms{km\,s$^{-1}$}
\def\deg{\hbox{$^\circ$}}

\def\fdg{\hbox{$.\!\!^\circ$}}
\def\farcm{\hbox{$.\mkern-4mu^\prime$}}

\begin{document}

\title{High--resolution imaging of compact high--velocity clouds}
\titlerunning{High--resolutions imaging of CHVCs}

\author{
  V. de\,Heij\inst{1},
  R. Braun\inst{2},
  and W.\,B. Burton\inst{1,3}
}

\institute{
  Sterrewacht Leiden,
    P.O. Box 9513,
    2300 RA Leiden,
    The Netherlands \and
  Netherlands Foundation for Research in Astronomy,
    P.O. Box 2,
    7990 AA Dwingeloo,
    The Netherlands \and
   National Radio Astronomy Observatory, 520 Edgemont Road,
Charlottesville, Virginia 22903, U.S.A.
}

\date{Received mmddyy/ accepted mmddyy}

\offprints{R. Braun\\
\email{rbraun@astron.nl}}

\abstract{ We have imaged five compact high--velocity clouds in \hi
with arcmin angular resolution and \kms~spectral resolution using the
Westerbork Synthesis Radio Telescope.  These CHVCs have a
characteristic morphology, consisting of one or more quiescent,
low--dispersion compact cores embedded in a diffuse warm halo.  The
compact cores can be unambiguously identified with the cool neutral
medium of condensed atomic hydrogen, since their linewidths are
significantly narrower than the thermal linewidth of the warm neutral
medium. Because of the limited sensitivity to diffuse emission inherent
to interferometric data, the warm medium is not directly detected in
the WSRT observations.  Supplementary total--power data, which is fully
sensitive to both the cool and warm components of H\,{\sc i}, is
available for comparison for all the sources, albeit with angular
resolutions that vary from 3$^\prime$ to 36$^\prime$. The fractional
\hi flux in compact CNM components varies from 4\% to 16\% in our
sample.  All objects have at least one local peak in the CNM column
density which exceeds about $10^{19}\rm\;cm^{-2}$ when observed with
arcmin resolution. It is plausible that a peak column density of
1--2$\times10^{19}\rm\;cm^{-2}$ is a prerequisite for the long--term
survival of these sources.  One object in our sample,
CHVC\,120$-$20$-$443 (Davies' cloud), lies in close projected proximity
to the disk of M\,31. This object is characterized by exceptionally
broad linewidths in its CNM concentrations, more than 5 times greater
than the median value found in the 13 CHVCs studied to date at
comparable resolution. These CNM concentrations lie in an arc on the
edge of the source facing the M\,31 disk. The diffuse \hi component of
this source, seen in total--power data from the NRAO 140--foot
telescope, has a positional offset in the direction of the M\,31
disk. All of these attributes suggest that CHVC\,120$-$20$-$443 is in a
different evolutionary state than most of the other CHVCs which have
been studied. Similarly broad CNM linewidths have only been detected in
one other cloud, CHVC\,110.6$-$07.0$-$466 (Wakker \& Schwarz
\cite{wakker91b}) which also lies in the Local Group barycenter
direction and has the most extreme radial velocity known.  A distinct
possibility for Davies' cloud seems to be physical interaction of some
type with M\,31. The most likely form of this interaction might be the
ram--pressure or tidal--stripping by either one of M\,31's visible
dwarf companions, M\,32 or NGC\,205, or else by a dark companion with
an associated \hi condensation. The compact objects located in the
direction of the Local Group barycenter have an important role to play
in constraining the Local Group hypothesis for the deployment of CHVCs.
 }

\maketitle

\section{Introduction}
Although high--velocity clouds have been studied extensively since
their discovery in~1963 by Muller \ea~(\cite{muller63}), there is still
no consensus on the origin and physical properties of these objects.
The clouds, for which most of the observations have been done in the
\hi 21--cm emission line, have velocities in excess of those allowed by
Galactic rotation.  Most of the physical properties like size, mass,
and gas density depend sensitively on the distances of the clouds:
these distances are still unknown, except in a few cases.  The
Magellanic Stream represents tidal debris originating in the
gravitational interaction of the Large and Small Magellanic Clouds with
our Galaxy (see Putman \&~Gibson, \cite{putman99}).  The Stream is
therefore likely located at a distance of about 50 kpc.  Other
high--velocity features with constrained distances are a few large
complexes, extending over some tens of square degrees.  One of these,
Complex~A, has been found from absorption--line observations (van
Woerden et al.  \cite{woerden99}, Wakker \cite{wakker01}) to lie within
the distance range of $8 < d < 10\rm\;kpc$.  Wakker \&~van Woerden
(\cite{wakker97}) and Wakker et al.  (\cite{wakker99}) have given
recent reviews of the high--velocity cloud phenomenon.

During the past several years, there has been a renewed interest in the
possibility that many high--velocity clouds are scattered throughout
the Local Group. This hypothesis has been considered by many authors
over the past decades: although some of the earlier references now
appear somewat outdated, several early studies seem to have been particularly
presentient: these include the work of Verschuur (\cite{verschuur}),
who discussed high--velocity clouds as protogalactic material scattered
throughout the Local Group, and the work of Eichler (\cite{eichler})
and Einasto et al.  (\cite{einasto}), who viewed high--velocity clouds
as carriers of dark matter, also scattered throughout the Local Group
and available for merger with the larger systems.

Cosmological simulations intended to represent the evolution of the
Local Group now predict a much higher number of dark--matter satellites
around our Galaxy and Andromeda than the number of observed dwarf
galaxies (Klypin et al.  \cite{klypin99}, Moore et al.
\cite{moore99}).  Although there are several possible solutions to this
problem, one is that the missing dark matter satellites should not only
be identified with dwarf galaxies, but also with the high--velocity
clouds.  These objects would have a very low star--formation rate,
consistent with the non--detection of stars or of emission from dust or
molecules associated with pre--stellar conditions.  Whereas Blitz et
al. (\cite{blitz99}) considered the properties of a general
high--velocity cloud catalog in search of evidence for this hypothesis,
Braun \&~Burton (\cite{braun99}) restricted their study to the compact
and isolated ones, the so-called CHVCs.  These objects are isolated in
the sense that they are not connected to extended emission features at
a level of \NH$ = 1.5\times10^{18}$\,cm$^{-2}$.  Such isolated objects
turn out to be very compact, having a median angular size of less than
$1^\circ$.  The signature of these small and compact clouds in the
Leiden/Dwingeloo survey (LDS, Hartmann \&~Burton \cite{hartmann97}) is
indistinguishable from that of a nearby dwarf galaxy.  If the
high--velocity clouds are the baryonic counterparts of low--mass
dark--matter halos, then the subset of compact and isolated objects
would be the most likely candidates for clouds at substantial
distances, as yet undistorted by tidal-- and ram--pressure stripping.

The visual search for CHVCs of Braun \&~Burton (\cite{braun99}) in the
LDS data has been extended by de\,Heij et al. (\cite{deheij02a}), with
a fully automated algorithm.  The same algorithm was used to isolate
the CHVC population in the southern hemisphere from the HIPASS data by
Putman et al. (\cite{putman02}).  The velocity dispersion of these
compact and isolated clouds is the lowest in the Local Group Standard
of rest system, lending some support to the idea that they are located
throughout the Local Group.  More importantly, self--consistent
modeling of \hi bound to a dark--matter mini--halo population in the
Local Group potential carried out by de\,Heij et al. (\cite{deheij02b})
gives support to this scenario. Critical aspects of this
modeling are the realistic treatment of the effects of foreground
obscuration by the \hi of our Galaxy, and the account taken of the
limited resolution and sensitivity of the existing survey data.

Due to its limited spatial resolution of~$36^\prime$ FWHM, the
Leiden/Dwingeloo survey is not an ideal basis for the study of the
internal \hi properties of the compact, high--velocity clouds.  Braun
\&~Burton (\cite{braun00}) obtained high--resolution WSRT observations
of six of these clouds. Other than the work by Braun \& Burton and that
reported here, only two CHVCs had previously been imaged at high
resolution, by Wakker \& Schwarz (\cite{wakker91b}).  The synthesis
observations reveal a characteristic morphology in which one or more
compact cores are embedded in a diffuse halo, confirming the results
from single--dish work done on large telescopes at moderately high
angular resolution, notably in the earlier work done on the NRAO
300--foot telescope, whose FWHM beam subtended 10 arcminutes at
$\lambda21$ cm, by Giovanelli et al. (\cite{giovanelli}).  The narrow
line widths characteristic of most core components seen at arcminute
resolution in the syntheis data allow unambiguous identification of
these with the cool condensed phase of HI, the CNM, with kinematic
temperatures near~$100\rm\;K$.  One of the CHVCs observed by Braun
\&~Burton (\cite{braun00}), CHVC\,125+41$-$207, showed several opaque
clumps with some of the narrowest \hi emission lines ever observed,
with intrinsic FWHM of no more than~$2\rm\;km\;s^{-1}$ and brightness
temperature of $75\rm\;K$.  From a comparison of column and volume
density for this object, Braun \&~Burton estimate a distance in the
range 0.5 to 1\,Mpc.  In addition, several of the compact cores show
systematic velocity gradients along the major axis of their elliptical
extent. Some of these are well--fit by circular rotation in a flattened
disk system. The apparent rotation velocities imply dark--matter masses
of about 10$^8$~M$_\odot$ and dark--to--visible mass ratios of
10\,--\,50 or more.  The cores of the multi--core objects show relative
velocities as large as~$70\rm\;km\;s^{-1}$ on 30~arcmin scales, also
implying either an extremely short dynamically lifetime or a high
dark--to--visible mass ratio.

In this paper, we extend the high--resolution study of CHVCs by imaging
an additional five clouds with the WSRT.  Our discussion is organized
as follows.  We begin by describing the method of sample selection in
\S\,\ref{sec:sample}, proceed with a description of the newly acquired
observations in \S\,\ref{sec:observations}, continue with a
presentation of the images in \S\,\ref{sec:presentation}, and conclude
with discussion of our results in \S\,\ref{sec:discussion}.

\section{Sample selection}
\label{sec:sample}

The sample was drawn from the CHVC catalog of de\,Heij et al.
(\cite{deheij02a}), which extends the original CHVC catalog produced by
Braun \&~Burton (\cite{braun99}).  Both catalogs are based on
candidates extracted from the Leiden/Dwingeloo survey, after obtaining
independent confirming data for each object. Braun \&~Burton
(\cite{braun00}) selected a sample of six CHVCs for high--resolution
WSRT imaging that spanned as wide a range in object parameters as
possible. In particular, the selected sources varied in \hi linewidth
from as little as 6 \kms~to as much as 95~\kms~FWHM, while the median
CHVC linewidth is about 30~\kms.  The current sample of five additional
objects was chosen to supplement this earlier one by targeting CHVCs
with a relatively narrow velocity width and a moderately high peak
brightness. This selection was motivated by the hope of detecting more
examples of the extremely compact, high--column--density clumps found
by Braun \& Burton in CHVC\,125+41$-$207. Resolved detection of such
clumps allows meaningful distance constraints to be placed on the
object.

Table~\ref{table:LDS} lists the basic properties of the CHVCs selected
for WSRT imaging and discussed here. The velocity FWHM refers to the
LDS spectrum with the peak brightness. The tabulated integrated fluxes
also were determined from the LDS data (de\,Heij et al.
\cite{deheij02a}), except in the case of CHVC\,120$-$20$-$443, which
extends in velocity beyond the LDS coverage. In this case, the Green
Bank 140--foot observations described below were used.
CHVC\,120$-$20$-$443 is especially interesting in that it is located
only $2^\circ$~from the nuclear position of M\,31, and has an extreme
radial velocity exhibited by only a few CHVCs, all of which lie in the
same general region, near the direction of the barycenter of the Local
Group. The proximity to M\,31 has previously led to speculation
regarding a possible association with that galaxy (Davies
\cite{davies75}).  CHVC\,186$+$19$-$114 has recently been mapped with
the Arecibo telescope with a spatial resolution of~$3^\prime$
(Burton~et al. \cite{burton01}).

\section{Observations}
\label{sec:observations}

Observations of the five CHVC fields were obtained with the WSRT during
July and August, 1999. The CHVCs at southern declinations could not be
observed with complete 12--hour tracks due to elevation
limitations. Although complete 12--hour tracks were scheduled for the
sources at northern declinations these were not completely
successful. The actual hour--angle coverage obtained for each source is
indicated in Table~\ref{table:WSRT}, together with the nominal
east-west separation of telescopes RT9 and RTA. Program observations
were preceded and followed by observations of one of the calibration
sources 3C48, 3C286, or 1938$-$155.  At the time of the observations
all 14~telescopes of the array were equipped with the upgraded MFFE
receivers.  These receivers have a system temperature of about~27~K in
the 1150 to 1850~MHz band.  The correlator was configured to provide
256~uniformly weighted spectral channels in two linear polarizations
across 2.5~MHz centered on the $V_{\rm LSR}$ velocity of each source.
The effective velocity resolution was 1.2~times the channel spacing,
which was 0.5 \kms~for CHVC\,148$-$82$-$258 and 1.0 \kms~for the other
sources.

Standard gain and bandpass calibrations were carried out after editing
the data for incidental interference and shadowing.  Self--calibration
utilizing continuum sources in the target fields has been used, where
necessary, to further calibrate the gains. This was particularly
required for the sources at negative declinations, which were observed
at relatively low elevations.  An image made from the average of the
emission--free spectral channels from each field provided a CLEAN
component model of the continuum emission.  This model was subtracted
directly from the visibility data.  The block of spectral channels
containing line emission was imaged with a visibility--based CLEAN
deconvolution proceeding down to a flux level of twice the rms noise
level.  Uniform weighting of the visibility data was employed together
with a Gaussian taper decreasing to 30\% amplitude at a projected
baseline of 1.25\,k$\lambda$.  The corresponding spatial resolution was
about 60~arcsec.  The velocity axis has been smoothed with a Gaussian
with a FWHM of~$1\rm\;km\;s^{-1}$ for CHVC\,148$-$82$-$258, and of
$2\rm\;km\;s^{-1}$ for the other sources.  Given the low observed
brightness of the sources, the application of the spatial taper and
velocity smoothing were required to get a usefully high
signal--to--noise ratio.  In a few cases, some residual continuum
emission was still present in the data cubes.  In those cases, several
spectral channels from both edges of the cube were averaged together
and subtracted from the entire cube.

The typical rms noise level in the deconvolved WSRT cubes listed in
Table~\ref{table:WSRT} was between 2.0 and 4.0~mJy per beam per
spectral channel, with the northern--declination cubes generally
superior in this respect to the southern ones.  The corresponding
brightness sensitivities are also listed in the Table~\ref{table:WSRT}.
(Flux per beam and brightness temperature are related as usual by $S =
2k_{\rm B} T_{\rm B} \Omega_{\rm B} / \lambda^2$, or $S_{\rm mJy/Beam}
= 0.65\;\Omega_{\rm as} T_{\rm B} / \lambda_{\rm cm}^2$, where
$\Omega_{\rm as}$ is the beam area in arcsec$^2$.)  Expressed as an
optically thin HI column density, the sensitivity corresponds to about
$0.4\times10^{18}\rm\;cm^{-2}$, for emission which fills the beam and
which extends over a single velocity channel of 2~\kms~width. Since
diffuse \hi in the halo component has a minimum observed linewidth of
about 24~\kms~FWHM, the more relevant column--density sensitivity is a
factor of $\sqrt12$ higher over this larger linewidth.

Moment images of zero, first, and second order were generated from each
cube, after employing a blanking criterion for inclusion of each pixel
in the weighted sum.  This involved demanding a brightness in excess of
about $2\sigma$ after smoothing the cube by an additional factor of
three, both spatially and in velocity.  Images of integrated emission
were corrected for the primary--beam response of the WSRT instrument,
which is well approximated, at 1420~MHz, by a circular Gaussian with
2110~arcsec FWHM.

By their nature, interferometers are insensitive to diffuse emission
more extended than about 1/$B_{\rm min}$ radians, for a minimum
baseline, $B_{\rm min}$, expressed in wavelengths.  The ratios between
the fluxes as measured with single--dish total--power observations and
the WSRT data clearly show that not all the flux is recovered; the
percentage of recovered flux is indicated in the last column of
Table~\ref{table:WSRT}. In general, only the narrow linewidth cores are
detected in the WSRT data. To compensate for this shortcoming, the WSRT
data are compared here with total--power data for the individual
sources.  For CHVC\,186$+$19$-$114, the total--power data are those
obtained from Arecibo observations made by Burton et
al. (\cite{burton01}); for CHVC\,148$-$82$-$258 and
CHVC\,358$+$12$-$137, HIPASS data were used, as both of these sources
lie in the zone of overlap between the LDS and the HIPASS material and
were also entered in the CHVC catalog of Putman et al.
(\cite{putman02}); for CHVC\,120$-$20$-$443, observations were made
using the 140--foot telescope of the NRAO in Green Bank; for
CHVC\,129$+$15$-$295, only LDS data were used.  The Arecibo
observations have a spatial resolution of about $3^\prime$; the HIPASS
observations (Barnes et al.  \cite{barnes01}), fully Nyquist sampled
with the 64--m Parkes telescope, have a spatial resolution of
$15\farcm5$~FWHM; the Green Bank 140--foot telescope had a beam size of
21$^\prime$~FWHM; the Dwingeloo 25--m telescope had a beam size of
$36^\prime$~FWHM.

The observations made with the NRAO 140--foot telescope of
CHVC\,120$-$20$-$443 were carried out during two runs, in
November/December, 1996, and in September, 1997, as part of a larger
program tracing anomalous--velocity \hi within some 10 degrees of
M\,31.  The observations were made in frequency--switching mode
(switching up 5 MHz), with a bandwidth of 5 MHz. Spectra were taken on
a 10--arcminute grid; the FWHM beam of the 140--foot antenna is $21'$
at $\lambda 21$\,cm. The spectral coverage extended from $V_{\rm
LSR}=-700$ \kms~to $+300$ \kms.  On--source integration times were 40
seconds. Conversion from antenna temperatures to brightness
temperatures followed from regular observations of the primary standard
field S8 and use of the conversion factors of Williams
(\cite{williams}).

\begin{table*}[b]
\caption{ Basic properties of the sample of CHVCs imaged with the WSRT,
based on data from the Leiden/Dwingeloo survey and additional
observations. The velocity FWHM pertains to the LDS spectrum with the
indicated peak brightness. The flux density of CHVC\,120$-$20$-$443
refers to observations made with the Green Bank 140--foot telescope.}
\renewcommand{\arraystretch}{1.10}
\begin{tabular}{lcclccc}
\hline
Object                     & {RA (2000)}
                         & {Dec (2000)}
                         & {LDS structure}
                         & {$T_{\rm max}$}
                         & {FWHM}
                         & {total flux} \\
CHVC $lll\pm bb \pm vvv$ & {$\rm(h\  m)$}
                         & {$\rm(^\circ\ ^\prime)$}
                         & {$\rm(a \times b\;@\;PA)$}
                         & {(K)}
                         & {(\kms)}
                         & {(Jy \kms)} \\
\hline
CHVC 120$-$20$-$443 & 00 38.2 & $+$42 28 & $0.^\circ4 \times
0.^\circ4$          & 0.29 & 18 &   95 \\
CHVC 129$+$15$-$295 & 02 33.2 & $+$76 40 & $0.^\circ8 \times
0.^\circ8$          & 0.44 & 18 &  120 \\
CHVC 148$-$82$-$258 & 01 05.0 & $-$20 16 & $0.^\circ8 \times
0.^\circ8$          & 0.47 & 20 &  140 \\
CHVC 186$+$19$-$114 & 07 16.9 & $+$31 46 & $0.^\circ9 \times
0.^\circ6\;@\;{-20^\circ}$ & 1.03 & 20 &  177 \\
CHVC 358$+$12$-$137 & 16 55.3 & $-$23 33 & $0.^\circ8 \times
0.^\circ8$          & 0.56 & 19 &  112 \\
\hline
\end{tabular}
\label{table:LDS}
\end{table*}

\begin{table*}[b]
\caption{ Parameters of the WSRT observations and some measured CHVC
    properties.
}
\renewcommand{\arraystretch}{1.10}
\begin{tabular}{llcccccc}
\hline
Object                     & {HA range}
                         & {$B_{\rm min}$}
                         & {resolution}
                         & {\sc rms}
                         & {\sc rms}
                         & {detected flux}
                         & {percentage}\\
CHVC $lll\pm bb \pm vvv$ & {~~(h$\rightarrow$h )}
                         & {$\rm(m)$}
& ($a''\times b''$ @ PA$^\circ$ $\times$ \kms)
                         & {$\rm(mJy/beam)$}
                         & {$\rm(K)$}
                         & { (Jy \kms)}
                         & {detected}\\
\hline
CHVC 120$-$20$-$443 & $-6 \rightarrow +1$  & 63 &
$198 \times 82\ @ -30\deg \times $ 2.1 & 2.7 & 0.10 & 8 & 8\% \\
CHVC 129$+$15$-$295 & $-4 \rightarrow +6$  & 54 &
$135 \times 98\ @ -19\deg \times $ 2.1 & 2.8 & 0.13 &  17 & 14\% \\
CHVC 148$-$82$-$258 & $-6 \rightarrow +3.5$  & 48 &
$514 \times 92\ @ \ -3\deg \times $ 1.0 & 4.4 & 0.06 &   6 & 4\% \\
CHVC 186$+$19$-$114 & $-3 \rightarrow +2.5$  & 54 &
$130 \times 87\ @ +28\deg \times $ 2.1 & 3.1 & 0.17 &  28 & 16\% \\
CHVC 358$+$12$-$137 & $-2 \rightarrow +2$  & 96 &
$468 \times 72\ @ -19\deg \times $ 2.1 & 3.3 & 0.06 &   4 & 4\% \\
\hline
\end{tabular}
\label{table:WSRT}
\end{table*}

\section{High--resolution images}
\label{sec:presentation}

\subsection{CHVC 120$-$20$-$443}
Discovered by Davies (\cite{davies75}), CHVC 120$-$20$-$443 is
especially interesting given its projected proximity to M\,31 and its
extreme velocity. The object is centered only about $2^\circ$ north of
the M\,31 nucleus and lies directly adjacent to the north--eastern disk
of that system, as shown in Fig.~\ref{fig:h120a}, which displays
integrated \hi data obtained with the NRAO 140--foot telescope.  The
moment map on the left side of this figure represents \NH~in the
velocity range $-490 < V_{\rm LSR} < -160$ \kms, and so includes much
of the full extent of M\,31; the moment map on the right shows \NH~from
the restricted range of velocities, $-470 < V_{\rm LSR} < -420$ \kms,
over which the CHVC itself contributes \hi emission.  There is no sign
of a bridge of \hi connecting the CHVC and M\,31 at the sensitivity of
these data, but there are suggestions of a physical influence of M\,31
on the cloud in other aspects of the observations which we discuss
below.
   
An overview of the WSRT results is given in Fig.~\ref{fig:h120b}. The
highest column densities in this CHVC are concentrated in a
semi--circular rim along the eastern periphery -- i.e. in the direction
of the M\,31 disk.  The object is characterized by rather disjoint
internal kinematics. Mean line--of--sight velocities vary over a range
of some 30~\kms, but do not do so smoothly. The line--of--sight
velocities are almost bi--modally distributed, with a more circularly
symmetric component near $V_{\rm LSR}=-$455~\kms, in addition to the
eastern rim feature centered near $-$440~\kms. The eastern rim of this
object is also remarkable for the broad velocity widths seen
there. Velocity dispersions as high as 10~\kms~are observed.  The broad
linewidths of this feature can be seen in the individual spectra of
Fig.~\ref{fig:h120c}; Gaussian fits to the spectra are listed in
Table~\ref{table:gauss}. None of the other objects presented here show
comparably broad linewidths in the cool cores detected with synthesis
imaging. In the earlier WSRT sample of Braun \& Burton (\cite{braun00})
such broad linewidths were only seen in systems that showed indications
of line--of--sight overlap of multiple distinct velocity systems. In
this case, however, the broad linewidths appear to be intrinsic to the
feature, or perhaps related to the bimodal velocity distribution noted
above.

The high--resolution WSRT channel maps are overlaid in
Fig.~\ref{fig:h120d} on the total--power data from the 140--foot
telescope. The diffuse \hi detected in the total--power data is
significantly offset toward the southeast from the core components seen
at high resolution.

\subsection{CHVC 129$+$15$-$295}

The extended environment of CHVC\,129$+$15$-$295 is illustrated in
Fig.~\ref{fig:h129a}, which shows on the left an image of
velocity--integrated \hi extracted from the Leiden/Dwingeloo
survey. The CHVC is completely isolated in position as well as in
velocity, down to the $3\sigma$ noise limit of the LDS, corresponding
to less than 1.5$\times$10$^{18}$\,cm$^{-2}$.  This object corresponds
to entry \#231 in the general catalog of Wakker \& van Woerden
(\cite{wakker91}), to entry \#31 in CHVC catalog of Braun \& Burton
(\cite{braun99}), and to \#72 in the CHVC catalog of de\,Heij et al.
(\cite{deheij02a}). The nearest high--velocity cloud complexes, as
cataloged by Wakker \& van Woerden (\cite{wakker91}, are Complex~A,
which extends to $(l=130^\circ,\,b=+22^\circ)$, and Complex~H located
around $(l=130^\circ,\,b=+5^\circ)$.  But the velocity differences with
respect to both of these complexes amount to more than
$70\rm\;km\;s^{-1}$: a physically--relevant relation of the CHVC to
either one of the complexes is not demonstrated.  The WSRT moment
images for this field are shown in Fig.~\ref{fig:h129b}.  The cloud
core has the shape of an inverted V--shaped filament which is brighter
on the west side than on the east. Peak column densities reach about
5$\times$10$^{19}$\,cm$^{-2}$.  The WSRT moment images show no
significant variation in the mean velocity or in the velocity
width. Representative WSRT spectra, shown on the righthand side of
Fig.~\ref{fig:h129a}, are all centered near $-305\rm\;km\;s^{-1}$, with
a FWHM of about 5~\kms.  The results of one--component Gaussian fits to
the spectra are tabulated in Table~\ref{table:gauss}. The maximum
velocity dispersion nowhere exceeds~$4\rm\;km\;s^{-1}$, and is
typically much less. The narrow widths imply both that the temperatures
are low and that the object is kinematically quiet.

\subsection{CHVC 186$+$19$-$114}

CHVC\,186$+$19$-$114, with a peak brightness temperature of 1.1 K at
half-degree angular resolution, is one of the brighter objects in the
de\,Heij et al. (\cite{deheij02a}) catalog of CHVCs. This object
corresponds to entry \#215 in the general catalog of Wakker \& van
Woerden (\cite{wakker91}), to entry \#44 in CHVC catalog of Braun \&
Burton (\cite{braun99}), and to \#92 in the catalog of de\,Heij et
al. A velocity--integrated intensity map, based on the Leiden/Dwingeloo
survey and shown in lefthand panel of Fig.~\ref{fig:h186a}, does not
reveal much detail, but illustrates well the isolated nature of this
source.  An overview of the WSRT data for CHVC 186$+$19$-$114 is given
in Fig.~\ref{fig:h186b}.  The velocity--integrated map shows an
ellipsoidal structure with the highest detected column densities along
both the eastern and northern edges as well as a single high--contrast
clump at ($\alpha,\,\delta$)=$(07^{\rm h}17^{\rm m}18^{\rm
s},\,31\deg33'36'')$, which reaches a peak column density of about
1.5$\times$10$^{20}$\,cm$^{-2}$.

The core/halo morphology of this CHVC is seen by comparing the Arecibo
observations of Burton et al. (\cite{burton01}) with the current WSRT
material.  An overlay of the Arecibo and WSRT observations, which
illustrates cloud structures down to a spatial resolution of about
$1\times 2\rm\;arcmin$, is shown in Fig.~\ref{fig:h186d}. Although
there is no obvious large--scale gradient apparent in the WSRT velocity
field, the sequence of Arecibo channel maps shown in
Fig.~\ref{fig:h186d} does indicate a clear gradient amounting to some
20~\kms~over 50~arcmin, extending from $-$125~\kms~in the northeast to
$-$105~\kms~in the southwest. The highest velocity dispersions are seen
toward the clump noted above. The doubly--peaked spectrum in this
direction, shown on the righthand side of Fig.~\ref{fig:h186a}, is
suggestive of velocity splitting of about 7~\kms. The parameters of the
best--fitting combination of two Gaussian components are listed in
Table~\ref{table:gauss}. This decomposition consists of a relatively
narrow component centered at $-122\rm\;km\;s^{-1}$, and a somewhat
broader component centered at $-115\rm\;km\;s^{-1}$. Gaussian
decompositions are notoriously non--unique under many common
circumstances, and often completely unphysical, so no undue
significance should be attached to these specific values.  A comparison
with the Arecibo spectra, indicated by the dashed lines in the lefthand
panel of Fig.~\ref{fig:h186a}, shows that the Arecibo data can be
described by a single broad component centered at an intermediate
velocity, $-118\rm\;km\;s^{-1}$, with no enhancement at
$-122\rm\;km\;s^{-1}$.  A second compact clump seen in the WSRT data at
($\alpha,\,\delta)=(07^{\rm h}17^{\rm m}25^{\rm s},\,31\deg43'12'')$
displays a similar effect. The narrow WSRT profile is centered at
$-122\rm\;km\;s^{-1}$, while the wider Arecibo profile is centered at
$-119\rm\;km\;s^{-1}$.  The effects of different angular resolution and
(lack of) sensitivity to the most diffuse structures are seen to
produce substantial differences in the spectra.

\begin{table}
\caption{ Gaussian fits to the brightness temperature spectra shown in
the figures.}
\begin{tabular}{rlrcc}
\hline
\multicolumn{1}{c}{RA (2000)}          &
\multicolumn{1}{c}{Dec (2000)}          &
\multicolumn{1}{c}{$T_{\rm peak}$}     &
\multicolumn{1}{c}{$V_{\rm LSR}$}      &
\multicolumn{1}{c}{FWHM}               \\
                                    &  &
\multicolumn{1}{c}{$(\rm K)$}          &
\multicolumn{1}{c}{$(\rm km\;s^{-1})$} &
\multicolumn{1}{c}{$(\rm km\;s^{-1})$} \\
\hline

\multicolumn{5}{c}{CHVC\,120$-$20$-$443} \\
\hline
$\rm00^h\,37^m\,26^s$ & $+42^\circ\,15'13''$
                      &           0.94   &           $-$448 & 16.0 \\
$\rm00^h\,38^m\,15^s$ & $+42^\circ\,14'16''$
                      &           0.67   &           $-$444 & 29.7 \\
$\rm00^h\,38^m\,23^s$ & $+42^\circ\,19'12''$
                      &           0.56   &           $-$441 & 25.8 \\
$\rm00^h\,38^m\,18^s$ & $+42^\circ\,26'11''$
                      &           0.40   &           $-$442 & 24.1 \\
$\rm00^h\,37^m\,21^s$ & $+42^\circ\,34'10''$
                      &           0.64   &           $-$437 & 18.5 \\
$\rm00^h\,37^m\,59^s$ & $+42^\circ\,29'41''$
                      &           0.41   &           $-$456 & 25.0 \\
\hline

\multicolumn{5}{c}{CHVC\,129$+$15$-$295} \\
\hline
$\rm02^h\,31^m\,21^s$ & $+76^\circ\,29'56''$
                      &            3.5   &           $-$305 &  5.8 \\
$\rm02^h\,30^m\,09^s$ & $+76^\circ\,29'42''$
                      &            2.7   &           $-$306 &  7.3 \\
$\rm02^h\,31^m\,48^s$ & $+76^\circ\,41'41''$
                      &            4.1   &           $-$308 &  5.3 \\
$\rm02^h\,31^m\,29^s$ & $+76^\circ\,47'50''$
                      &            4.5   &           $-$305 &  4.4 \\
$\rm02^h\,34^m\,29^s$ & $+76^\circ\,36'58''$
                      &            3.5   &           $-$305 &  5.3 \\
\hline
\multicolumn{5}{c}{CHVC\,148$-$82$-$258} \\
\hline
$\rm01^h\,04^m\,55^s$ & $-20^\circ\,11'40''$
                      &            0.63  &           $-$272 &  9.8 \\
$\rm01^h\,04^m\,56^s$ & $-20^\circ\,20'44''$
                      &            0.31  &           $-$271 &  7.7 \\
$\rm01^h\,03^m\,39^s$ & $-20^\circ\,17'01''$
                      &            0.30  &           $-$272 & 10.2 \\
\hline
\multicolumn{5}{c}{CHVC\,186$+$19$-$114, Arecibo} \\
\hline
$\rm07^h\,17^m\,18^s$ & $+31^\circ\,33'36''$
                      &            3.9   &           $-$118 & 13.3 \\
$\rm07^h\,17^m\,39^s$ & $+31^\circ\,41'28''$
                      &            5.6   &           $-$117 &  9.4 \\
$\rm07^h\,17^m\,25^s$ & $+31^\circ\,43'12''$
                      &            5.9   &           $-$119 & 10.4 \\
$\rm07^h\,16^m\,29^s$ & $+31^\circ\,49'34''$
                      &            4.0   &           $-$118 & 11.8 \\
$\rm07^h\,18^m\,11^s$ & $+31^\circ\,53'56''$
                      &            3.6   &           $-$119 & 10.4 \\
\hline

\multicolumn{5}{c}{CHVC\,186$+$19$-$114, WSRT} \\
\hline
$\rm07^h\,17^m\,18^s$ & $+31^\circ\,33'36''$
                      &            7.6   &           $-$122 &  4.8 \\
                  &   &            4.4   &           $-$115 &  5.8 \\
$\rm07^h\,17^m\,39^s$ & $+31^\circ\,41'28''$
                      &            7.0   &           $-$117 &  6.3 \\
$\rm07^h\,17^m\,25^s$ & $+31^\circ\,43'12''$
                      &            8.4   &           $-$122 &  3.9 \\
$\rm07^h\,16^m\,29^s$ & $+31^\circ\,49'34''$
                      &            4.7   &           $-$118 &  7.7 \\
$\rm07^h\,18^m\,11^s$ & $+31^\circ\,53'56''$
                      &            3.5   &           $-$119 &  7.7 \\
\hline
\multicolumn{5}{c}{CHVC\,358$+$12$-$137} \\
\hline
$\rm16^h\,56^m\,09^s$ & $-23^\circ\,32'53''$
                      &            0.27  &           $-$152 &  7.7 \\
                  &   &            0.40  &           $-$142 &  3.9 \\
$\rm16^h\,55^m\,24^s$ & $-23^\circ\,35'46''$
                      &            0.43  &           $-$138 &  8.7 \\
$\rm16^h\,55^m\,14^s$ & $-23^\circ\,25'35''$
                      &            0.31  &           $-$138 &  6.3 \\
$\rm16^h\,54^m\,25^s$ & $-23^\circ\,28'27''$
                      &            0.33  &           $-$143 &  7.3 \\
\hline
\end{tabular}
 \label{table:gauss}
\end{table}

\subsection{CHVC 148$-$82$-$258 and CHVC 358$+$12$-$137}

These two compact high--velocity clouds were discovered by Braun \&
Burton (\cite{braun99}) in the LDS material, and enter their catalog as
\#36 and \#66, respectively.  CHVC\,148$-$82$-$258 also corresponds to
entry \#67 in the CHVC catalog of de\,Heij (\cite{deheij02a}) and to
entry \#1545 in the Putman et al. (\cite{putman02}) HIPASS
catalog. CHVC\,358$+$12$-$137 corresponds to \#109 in de\,Heij
et al., and to \#2165 in Putman et al. The southerly declinations of
these two sources limited the image quality that could be obtained with
the WSRT array (which is sited at 54$^\circ$ north geographic
latitude). The limited (U,V) coverage listed in Table~\ref{table:WSRT}
resulted in a highly elongated synthesized beam, of about $1^\prime
\times 8^\prime$, and a high sidelobe level. Even after deconvolution,
the resulting image fidelity was not high, as judged by the
non--Gaussian character of the deconvolution residuals.

The WSRT results for these two objects are summarized in
Figs.~\ref{fig:h148a} and~\ref{fig:h358a}, respectively.  The figures
show the WSRT integrated \hi contours overlaid on the HIPASS
total--power data.  Only a few percent of the total flux has been
detected by the interferometer in these objects, as indicated in
Table~\ref{table:WSRT}.  Peak column densities are only about
1$\times$10$^{19}$\,cm$^{-2}$.  The locations of the compact structures
detected in the synthesis data are not coincident with the locations of
the brightest regions of the 15$^\prime$ resolution HIPASS data.
Representative spectra of both clouds are shown in the righthand panels
of Figs.~\ref{fig:h148a} and~\ref{fig:h358a}, respectively.  The
results of Gaussians fits to the spectra are given in
Table~\ref{table:gauss}.

The emission detected by the WSRT in CHVC\,148$-$82$-$258 is
concentrated in a single elongated clump, centered near
$-272\rm\;km\;s^{-1}$. Small differences in the centroid velocity, of
about 1~\kms, are seen at the various positions.

The WSRT observations of CHVC\,358$+$12$-$137 show four major clumps,
with centroid velocities ranging from~$-144$ to~$-138\rm\;km\;s^{-1}$.
The most easterly of these clumps, at ($\alpha,\,\delta)=(16^{\rm
h}56^{\rm m}09^{\rm s},\,23\deg32'53'')$, shows some indication for
line--splitting, although the signal--to--noise ratio is low.  At face
value this splitting amounts to some 10~\kms, as listed in
Table~\ref{table:gauss}.

\section{Discussion}
\label{sec:discussion}

\subsection{Cool and warm neutral media}

Wolfire~et al.\ (\cite{wolfire95}), following on the earlier treatments by
Field~et al. (\cite{field}) and Draine (\cite{draine}), have shown that
diffuse \hi clouds in thermodynamic equilibrium might have a two--phase
temperature structure.  The two components, a cool one (CNM) with
temperatures around~$100\rm\;K$, and a warm one (WNM) with temperatures
around~$10^4\rm\;K$, can coexist in pressure equilibrium for thermal
pressures, $P/k_B$, in the range of about 100 -- 2000~cm$^{-3}$\,K.  The
calculations presented in Wolfire et al.\ (\cite{wolfire95}) have been
supplemented with new ones appropriate for a low metallicity
population of CHVCs residing at significant distances in the Local
Group environment. The results of these new calculations are shown in
Fig.~13 of Braun \& Burton (\cite{braun00}), where equilibrium
solutions are given for clouds with a shielding column density of~1 and
$10\times10^{19}\rm\;cm^{-2}$, a metallicity of 0.1~solar, and a
dust--to--gas mass ratio of 0.1 times the value in the solar neighborhood.

The velocity FWHM of \hi clouds with kinetic temperatures of 100 K
and~$10^4$ K are 2.4 \kms~and 24 \kms, respectively. As shown in
Table~\ref{table:gauss}, the values observed in the
high--colum--density cores detected in the WSRT observations of our
sample vary between FWHM of about 4 and 30~\kms. No new example of
ultra--narrrow \hi lines was detected, such as the 2~\kms~FWHM features
seen in CHVC\,125+41$-$207 by Braun \& Burton (\cite{braun00}).  The
median linewidth of the material being discussed here amounts to about
6~\kms~FWHM, a width comparable to that seen at high resolution in both
HVC and CHVC studies (Wakker \& Schwarz \cite{wakker91}, Braun \&
Burton \cite{braun00}) as well as in nearby external spiral galaxies
(Braun \cite{braun97}). This is somewhat broader than expected for the
thermal linewidth of a 100~K gas, suggesting that either some form of
ordered or random internal motions are present, or that the available
resolution does not adequately account for line--of--sight blending of
separate components.  The alternative, namely that the typical kinetic
temperature is actually about 650~K, seems to be ruled out by
observations of \hi absorption in HVCs, as well as in the Galaxy and
external galaxies, which reveal spin temperatures of 50 to 175~K in all
cases (see Wakker et al.  \cite{wakker91c}, Braun \& Walterbos
\cite{braun92}, and Braun \cite{braun95}).

The large velocity widths, of 25 to 30~\kms~FWHM, found in the
condensations along the eastern rim of CHVC 120$-$20$-$443 are
difficult to interpret in this context. Of the thirteen CHVCs which
have currently been subjected to arcmin resolution synthesis imaging,
only one other instance of broad CNM linewidths has been observed,
namely in CHVC\,110.6$-$07.0$-$466 (Hulsbosch's cloud) as imaged by
Wakker \& Schwarz (\cite{wakker91b}). Several other cases of broad
widths are not relevant in this context, because they could be
unambiguously attributed to line--of--sight overlap of components at
different velocities.  As noted above in \S\,\ref{sec:presentation}, it
is not clear whether the linewidths in this feature are intrinsic or
whether the velocity field becomes systematically double--valued at
this location. We will return to this issue in a following subsection.

In general, it seems clear that it is predominantly the CNM which is
detected in the WSRT images for objects of $0\fdg5$ or more in
size. The smoothly distributed WNM can not be readily detected in the
synthesis data.  The fractional flux of CNM in the five objects studied
here varies from about 4\% to 16\%. This fraction varied from less than
1\% to more than 50\% in the Braun \& Burton (\cite{braun00}) sample,
which spanned a larger range of source properties. It is noteworthy
that in all CHVCs studied to date with high spatial resolution there
has been at least a marginal detection of the CNM. Every one of the
thirteen CHVCs studied to date has at least one local peak in the CNM
column density which exceeds about $10^{19}\rm\;cm^{-2}$ when observed
with arcmin resolution. The accompanying diffuse WNM halo reaches
comparable peak column densities, of about 1 or
2$\times10^{19}\rm\;cm^{-2}$, external to these peaks (Burton et
al. \cite{burton01}).  It is conceivable that a WNM halo column density
of 1 or 2$\times10^{19}\rm\;cm^{-2}$ is a prerequisite for the
long--term survival of these sources. It may be no coincidence that
Maloney (\cite{maloney93}) and Corbelli \& Salpeter (\cite{corbelli93})
estimated this value as the critical column density needed to prevent
complete ionization of the \hi when exposed to the estimated
extragalactic ionizing radiation field relevant for free--floating
objects in the Local Group.

\subsection{Velocity offsets of cool cores and warm halos}

In the case of CHVC\,186$+$19$-$114, it was possible to make a detailed
comparison of line profiles as measured in CNM cores using the WSRT
with the sum of the CNM and WNM emission detected in the 3--arcmin beam
of the Arecibo telescope. The CNM spectra show narrower intrinsic
widths as well as some local differences in the centroid velocity,
while the Arecibo spectra display broad--linewidth tails (consistent
with a 10$^4$~K thermal component) and much less dramatic variation in
the profile shape and centroid. Given the dominant role of the WNM,
accounting for about 84\% of the total \hi flux in this source, these
differences are not surprising. Although the WSRT and Arecibo velocity
centroids often agree, there are a few isolated locations where the CNM
component is offset from the total \hi centroid by a few \kms. If a
systematic velocity offset had been apparent between the WSRT and
Arecibo spectra, it might have been an indication for an external
perturbation of the source.

Br\"uns et al. (\cite{bruens01}), who have observed the interesting
object CHVC\,125$+$41$-$207 with the 100--m Effelsberg telescope, argue
that there is a systematic velocity offset between a narrow and broad
component of the \hi emission in that source. Their conclusion is based
on Gaussian decompositions of the slightly asymmetric line profiles in
the Effelsberg spectra. The decompositions result in two components;
one of about 5 \kms~and the other of 12 \kms~FWHM. It is difficult to
assess the physical relevance of these decomposition results, since at
large distances from the CHVC centroid a single Gaussian of about
20~\kms~is found to suffice in fitting the Effelsberg spectra well,
while within the CHVC centroid, the WSRT data of Braun \& Burton
(\cite{braun00}) for this object show non--Gaussian CNM line profiles
of only 2 to 4~\kms~FWHM.  Given the intrinsic non--uniqueness of
Gaussian decomposition when applied to non--Gaussian line profiles, it
seems questionable whether the 5 and 12 \kms~FWHM Gaussian--fit
components refer to physical systems at all.

If a systematic offset of the CNM and WNM velocities were present, then
this might indicate that the halo kinematics is perturbed by an
external force, which has not yet perturbed the central core of the
cloud.  The gravitational tidal field of either the Galaxy or M\,31 is a
candidate for such a differential force. Another possibility is the
ram--pressure exerted on the cloud as it moves through an external
medium.  Given the substantial differences in sound--crossing times of
the cores relative to the halo, a significant time delay in the
response might result.

\subsection{The particular interest of CHVC 120$-$20$-$443}

In his 1975 paper, Davies considered two possible interpretations of
this cloud.  Given its proximity to M\,31 on the sky, it might be
located at a comparable distance, with a projected separation of
only~$18\rm\;kpc$.  Since peak column densities are only a few times
$10^{19}\rm\;cm^{-2}$, internal star formation is unlikely: our
non--detection of stars in Palomar Sky Survey prints is no
surprise. With only the visible baryonic mass, Davies concluded that
the cloud is not gravitationally bound, and will double its size on a
time scale of~$2.4\times10^8$~years.  As an alternative possibility he
considered that the cloud might be related in some way to the
Magellanic Stream.  The closest approach of this feature to portions of
the Stream is, however, about 30$^\circ$ in angle and about 65~\kms~in
velocity, making such an association tenuous at best.  If the cloud
were a part of the Magellanic Stream, its distance might be about
$60\rm\;kpc$.  If there were no confining force except the
self--gravity of the cloud, it would double its size in approximately
$2\times10^7\rm\;years$.

Our high--resolution imaging of CHVC\,120$-$20$-$443 provides some
insights into the possible origin of this object. As noted in
\S\,\ref{sec:presentation}, the high--column--density cores in this
source are concentrated in a semi--circular rim along the eastern
periphery, in the direction of the M\,31 disk. Furthermore,
exceptionally broad linewidths, of 25 to 30~\kms~FWHM, are seen in this
rim feature, while enhanced linewidths, amounting to 15 to 20~\kms, are
seen throughout the source. Of the thirteen CHVCs studied to date with
arcmin resolution, only CHVC\,110.6$-$07.0$-$466 has shown comparably
broad linewidths in the CNM cores that are detected in interferometric
data.  Median linewidths in the CNM cores of CHVCs imaged by Wakker \&
Schwarz \cite{wakker91}, Braun \& Burton \cite{braun00}, and in this
paper are only 6~\kms. As noted previously, it is not yet clear whether
the broad linewidths are intrinsic, or due to a large-scale geometric
effect. One possibility might be a large physical extent along the
line--of--sight.  Another curious circumstance is the large spatial
offset of the brightest diffuse \hi detected in the Green Bank
140--foot data toward the southeast of the CNM rim, as seen in
Fig.~\ref{fig:h120d}. All of these observations suggest that
CHVC\,120$-$20$-$443 is in a different evolutionary state than the
other CHVCs which have been studied. Wakker \& Schwarz
(\cite{wakker91b}) suggest a similarly different evolutionary state for
CHVC\,110.6$-$07.0$-$466.  A distinct possibility seems to be a
physical interaction of some type with M\,31.

It is interesting to speculate how an observer in M\,31 would see
CHVC\,120$-$20$-$443 if it were at the relative
distance of 18~kpc.  Given the properties of the cloud, the M\,31
observer's perception of it could resemble the impression an
earth-based observer has of the HVC Complexes~A or~C. For an observer
located in the center of M\,31, the cloud would extend over
some~$30^\circ$ on the sky.  Lower limits to the peak column densities
that the observer would measure are determined by the ones measured in
the WSRT observations, which have values of a few times
$10^{19}\rm\;cm^{-2}$.  The WSRT observations show a filamentary
structure with several embedded higher--density clumps.  The relative
velocity of the object would be about 140~\kms, given the M\,31
systemic velocity of $-$300~\kms. In order for this velocity to
correspond to infall toward M\,31 the object would have to be located
beyond M\,31, rather than between M\,31 and the Galaxy.  From our
vantage point in the Galaxy, the HVC Complex~A extends over
about~$30^\circ$ on the sky, while Complex~C extends over
some~$70^\circ$. Both have radial velocities of about $-$100~\kms~in
the Galactic Standard of Rest frame, and peak column densities of about
$10^{19}\rm\;cm^{-2}$ as measured in the Leiden/Dwingeloo survey.
Concerning distances, we note that Complex~A is well constrained to lie
between 8 and 10~kpc (van Woerden et al. \cite{woerden99}, Wakker
\cite{wakker01}), while only a few lower limits are available for
Complex~C.  Although these clouds do not agree perfectly regarding
their observable HI properties, they resemble each other sufficiently
that it seems plausible to speculate about a similar physical origin.

Given the substantial projected distance of CHVC\,120$-$20$-$443 from
M\,31, an origin in a galactic fountain within that galaxy seems
unlikely.  In a galactic fountain, gas which is heated and ionised by
supernova explosions rises to higher z--height, either buoyantly or
driven by subsequent supernovae, where it finally condenses and returns
in free fall back toward the galactic disk (see Shapiro \& Field
\cite{shap76}, Bregman~\cite{bregman80}). Simulations carried out by
de\,Avillez (\cite{avillez00}) suggest that the height of this
condensation process is at most several kpc above the stellar disk.
CHVC\,120$-$20$-$443 is located substantially further away from the
stellar disk of M\,31.  Because the driving force of a galactic
fountain is provided by supernova explosions, which are concentrated in
OB--associations, it is remarkable that only one such cloud would be
seen.  The location of CHVC\,120$-$20$-$443 is also not correlated with
any region in M\,31 of particularly active star formation (see Pellet
et al. \cite{pellet}), making this scenario appear unlikely.

A tidal origin for CHVC\,120$-$20$-$443, related to either M\,32 or
NGC~205, is worth considering.  Ibata et al. (\cite{ibata01}) have
discovered a tidal stream of metal--rich stars extending several
degrees toward the south of M\,31.  They consider the dwarfs M\,32 or
NGC~205 as possibly responsible for the origin for the stream.  The
angular extent of the stellar stream toward the south is comparable to
the separation of CHVC\,120$-$20$-$443 from the center of M\,31 toward
the north. Together these systems might trace portions of the same
orbital path.  However, the measured radial velocity of the cloud is
difficult to reconcile with those of the dwarfs. Both dwarfs have
positive radial velocities with respect to~M\,31 ($+155\rm\;km\;s^{-1}$
in the case of M\,32 and $+59\rm\;km\;s^{-1}$ in the case of NGC~205)
whereas the high--velocity cloud has a negative relative velocity of
$-145\rm\;km\;s^{-1}$.  According to the distances listed in Mateo
(\cite{mateo98}), NGC~205 is located beyond M\,31.  Combined with its
positive velocity with respect to M\,31, it could be moving away from
its peri--center passage.  During closest approach, the gas could have
been stripped, either by ram--pressure stripping or by tidal
disruption. However, the deceleration of the gas by some
$200\rm\;km\;s^{-1}$ would need to be accounted for. Realistic
hydrodynamic simulations of such encounters might be illuminating.

Finally, the cloud could be part of a Local Group population of \hi
condensations within low--mass dark-matter halos, as described in the
Local Group deployment model of CHVCs (Blitz et al. \cite{blitz99},
Braun \& Burton \cite{braun99}). Analysis of the all--sky population of
CHVCs performed by de\,Heij et al. (\cite{deheij02b}) has resulted in a
self--consistent scenario whereby the observed CHVCs are part of a
power--law distribution in baryonic mass (with slope $-1.7$) coupled to
a steeper power--law (with slope $-$2) in dark mass. Only within the
\hi mass range of some 10$^{5.5}$ to 10$^7$ M$_\odot$ are the objects
stable against complete ionization by the intergalactic radiation field
on the one hand (at low mass), and stable to internal star formation on
the other (at high mass). The best--fitting simulated spatial
distributions are centered on each of the Galaxy and M\,31 with a
spatial Gaussian dispersion of some 150~kpc. The majority of currently
detected CHVCs belong to the relatively nearby swarm centered on the
Galaxy. Only a small fraction of the M\,31 sub--concentration of CHVCs
is predicted to have been bright enough for detection in the current
\hi surveys. At the distance of M\,31, CHVC\,120$-$20$-$443 has an \hi
mass of about 10$^7$ M$_\odot$, putting it at the high--mass end of the
distribution.  If the projected separation with respect to M\,31 is a
measure for its real distance, then the cloud is sufficiently close to
be strongly perturbed by the ram--pressure of its motion through a
gaseous halo around M\,31 (see de\,Heij et al.  \cite{deheij02b}). The
observed extreme CNM linewidths in this object, and the significant
displacement of the diffuse gas in the direction of M\,31 with respect
to the core components, may both be evidence for such an ongoing
perturbation.

Of all of the CHVCs extracted by de\,Heij et al. (\cite{deheij02a})
from the LDS together with those found in the HIPASS material by Putman
et al. (\cite{putman02}) and comprising an all--sky sample, only six
have a velocity more extreme than $|V_{\rm LSR}|=400$ \kms.  All of
these objects have negative velocities, and all lie at northern
declinations; they constitute the population of clouds often called
VHVCs.  Arguments that this kinematic envelope is not an artifact of
the observational parameters are given by de\,Heij et
al. (\cite{deheij02b}). (The most extreme positive--velocity CHVC is
the HIPASS object CHVC\,258.2$-$23.9$+$359; the most extreme
negative--velocity CHVC at southern declinations is
CHVC\,125.1$-$66.4$-$353.)  The most extreme--velocity CHVCs are the
following, using the designation given by de\,Heij et al. and, in
parenthesis, the entry numbers from the catalogs of Wakker \& van
Woerden (\cite{wakker91}), Braun \& Burton (\cite{braun99}), and
de\,Heij et al. (\cite{deheij02a}): CHVC\,103.4$-$40.1$-$414 (WW\#491,
deH\#57), CHVC\,107.7$-$29.7$-$429 (WW\#437, BB\#22, deH\#59),
CHVC\,108.3$-$21.2$-$402 (WW\#389, BB\#23, deH\#60),
CHVC\,110.6$-$07.0$-$466 (WW\#318, BB\#24, deH\#61),
CHVC\,113.7$-$10.6$-$441 (WW\#330, BB\#25, deH\#62), and Davies' cloud
CHVC\,120.2$-$20.0$-$444 (deH\#68).  These CHVCs cluster near the
direction of the barycenter of the Local Group, and are
characteristically faint and small: they are likely to play an
important role in the continuing discussion of the Local Group
hypothesis.

The simulations of the Local Group hypothesis reported by de\,Heij et
al.  (\cite{deheij02b}) support the prediction that a substantial
number of additional CHVCs at extreme velocities will be found in the
general direction of the Local Group barycenter, i.e. near M\,31, when
the sensitivity of the available \hi survey data is improved.  The
unusual properties of Davies' cloud may be revealed by other objects.
Two of the extreme--velocity objects (both discovered by Hulsbosch,
\cite{hulsbosch}), namely CHVC\,113.7$-$10.6$-$441 and
CHVC\,110.6$-$07.0$-$466, have been subject to synthesis imaging by
Wakker \& Schwarz (\cite{wakker91b}).  It is interesting to note that
Wakker \& Schwarz state that the properties of these CHVCs differ
considerably from the properties of the extended HVCs which they also
partly imaged. CHVC\,110.6$-$07.0$-$466 showed the same broad linewidth
properties as we have found here for Davies' cloud.  It is plausible
that the two objects have undergone a similar evolutionary
experience.

\subsection{Summary and conclusions}

We have imaged five CHVCs in \hi with arcmin angular resolution and
\kms~spectral resolution using the Westerbork Synthesis Radio
Telescope. These five images raise to 13 the number of CHVCs which have
been subject to synthesis mapping, including the two compact objects
studied by Wakker \& Schwarz (\cite{wakker91b}) and the six studied by
Braun \& Burton (\cite{braun00}). These objects have a characteristic
morphology, consisting of one or more quiescent, low--dispersion
compact cores embedded in a diffuse warm halo.  The compact cores can
be unambiguously identified with the cool neutral medium of condensed
atomic hydrogen, since their linewidths are significantly narrower than
the thermal linewidth of the warm neutral medium. Because of the
limited sensitivity to diffuse emission inherent to interferometric
data, the warm medium is not directly detected in the synthesis
observations discussed here.  Supplementary total--power data, which is
fully sensitive to both the cool and warm components of H\,{\sc i}, is
available for all sources for comparison, although with angular
resolutions that vary from 3$^\prime$ to 36$^\prime$. The fractional
\hi flux in compact CNM components varies from 4\% to 16\% in our
sample.  All objects have at least one local peak in the CNM column
density which exceeds about $10^{19}\rm\;cm^{-2}$ when observed with
arcmin resolution. The accompanying diffuse WNM halo reaches comparable
peak column densities, of about 1--2$\times10^{19}\rm\;cm^{-2}$,
external to these peaks (Burton et al. \cite{burton01}).  It is
conceivable that a WNM halo column density of
1--2$\times10^{19}\rm\;cm^{-2}$ is a prerequisite for the long--term
survival of these sources.

One object in our sample, CHVC\,120$-$20$-$443 (Davies' cloud), lies in
close projected proximity to the disk of M\,31. This object is
characterized by extremely broad linewidths in its CNM concentrations,
which are 5 to 6 times broader than the median value found in the 13
objects studied to date at comparable resolutions. The CNM
concentrations lie in an arc on the edge of the source facing the M\,31
disk. The diffuse \hi component of this source, seen in total--power
data, has a large positional offset in the direction of the M\,31
disk. All of these attributes suggest that CHVC\,120$-$20$-$443 is in a
very different evolutionary state than the other CHVCs which have been
studied, with the possible exception of CHVC\,110.6$-$07.0$-$466
(Hulsbosch's cloud), imaged by Wakker \& Schwarz (\cite{wakker91b}) and
shown to also have broad CNM clumps. A distinct possibility seems to be
a physical interaction of some type with M\,31. The most likely form of
this interaction might be ram--pressure or tidal--stripping by one of
M\,31's visible dwarf companions, M\,32 or NGC~205, or by a dark
companion with an associated \hi condensation.

\begin{acknowledgements}
The Westerbork Synthesis Radio Telescope is operated by the Netherlands
Foundation for Research in Astronomy under contract with the
Netherlands Organization for Scientific Research.  The National Radio
Astronomy Observatory is operated by Associated Universities, Inc.,
under contract with the US National Science Foundation.  The Arecibo
Observatory is part of the National Astronomy and Ionosphere Center,
which is operated by Cornell University under contract with the US
National Science Foundation. The Parkes telescope is part of the
Australia Telescope which is funded by the Commonwealth of Australia
for operation as a National Facility managed by CSIRO.
\end{acknowledgements}

\begin{figure*}
\begin{tabular}{cc}
\end{tabular}
\caption{Two images of integrated \hi emission illustrating the
environment of Davies' cloud, CHVC\,120$-$20$-$443.~~{\it left:} \hi
emission integrated over the velocity interval $-490 < V_{\rm LSR} <
-160$ \kms, a range which encompasses most of the emission from M\,31.
Contours of \hi column density are drawn at 4, 8, 12, 16, and
20$\times10^{18}$cm$^{-2}$. The grey--scale bar in both panels is
labelled in units of $10^{18}$\,cm$^{-2}$.  The disk of M\,31 is
directly adjacent to the CHVC although offset in velocity.  ~~{\it
right:} \hi emission integrated over the velocity interval $-470 <
V_{\rm LSR} < -420$ \kms, corresponding to the total velocity extent of
the CHVC.  Contours of \NH~are drawn at levels of $-$1, 1, 2, 4, 8, 12,
and 16$\times10^{18}$cm$^{-2}$. No bridge of emission is evident at
these velocities between the CHVC and M\,31, although the centroid of
the cloud emission is suggestively skewed towards M\,31.
}\label{fig:h120a}
\end{figure*}

\begin{figure*}
\begin{tabular}{ccc}
\end{tabular}
\caption{CHVC\,120$-$20$-$443 as imaged with the WSRT at resolutions of
$2\times1$~arcmin and $2\rm\;km\;s^{-1}$.~~{\it left:} Integrated \hi
with contours drawn at 2.5,\,5.0,\,7.5, and 10.0 K\,\kms, and a linear
grey--scale extending from~$-1$ to 12 K\,\kms.~~{\it middle:}
Intensity--weighted line--of--sight velocity, $V_{\rm LSR}$, with
contours drawn at $-458,\,-456,\,\ldots,\,-432$ \kms, and a linear
grey--scale extending from~$-460$ to~$-430$ \kms.~~{\it right:}
Distribution of the intensity--weighted velocity dispersion, with a
linear grey--scale extending from~$0$ to~$11$ \kms. The contour
corresponds to a dispersion of 5~\kms.  }\label{fig:h120b}
\end{figure*}

\begin{figure}
\caption{Brightness temperature WSRT spectra of CHVC\,120$-$20$-$443 at the
    indicated positions.
}\label{fig:h120c}
\end{figure}

\begin{figure*}
\begin{center}
\end{center}
\caption{Overlay of WSRT and Green Bank 140--foot channel maps for
CHVC\,120$-$20$-$443. The WSRT contours are drawn at 0.4, 0.6, 0.8, and
1.0 K; the data from the 140--foot telescope are represented as a
grey--scale extending from 0 to 250 milli--Kelvin.  Only about 8\% of
the total flux measured in the tota--power observations is recovered by
the interferometer.  }\label{fig:h120d}
\end{figure*}

\begin{figure*}
\begin{tabular}{cc}
\end{tabular}
\caption{~~{\it left:} Velocity--integrated intensity map of
CHVC\,129$+$15$-$295 derived from the Leiden/Dwingeloo survey.  The
range of integration extends from the average velocity minus the FWHM
to the average velocity plus the FWHM.  The single contour is drawn at
$N_{\rm HI}= 1.5\times10^{18}$\,cm$^{-2}$.~~{\it right:} Brightness
temperature spectra of CHVC\,129$+$15$-$295, measured with the WSRT at
the indicated positions.  }\label{fig:h129a}
\end{figure*}

\begin{figure*}
\begin{tabular}{ccc}
\end{tabular}
\caption{CHVC\,129$+$15$-$295 at $2\times1$~arcmin and
$2\rm\;km\;s^{-1}$ resolution.~~{\it left:} Apparent integrated HI in
$\rm K\;km\;s^{-1}$ with contours
at~$5,\,10,\,15,\,20\rm\;K\;km\;s^{-1}$ and a linear grey--scale
extending from~$-1$ to~$25\rm\;K\;km\;s^{-1}$.~~{\it middle}
Intensity--weighted line--of--sight velocity, $V_{\rm LSR}$, with
contours drawn at $-312,\,-310,\,\ldots,\,-298\rm\;km\;s^{-1}$ and a
linear grey--scale extending from~$-312$
to~$-297\rm\;km\;s^{-1}$.~~{\it right:} Distribution of the
intensity--weighted velocity dispersion, with a linear grey--scale
extending from~$0$ to~$5\rm\;km\;s^{-1}$. The contour corresponds to a
dispersion of 2~\kms.  }\label{fig:h129b}
\end{figure*}

\begin{figure*}
\begin{tabular}{cc}
\end{tabular}
\caption{~~{\it left:} Velocity--integrated intensity map of
CHVC\,186$+$19$-$114 derived from the Leiden/Dwingeloo survey.  The
range of integration extends from the average velocity minus the FWHM
to the average velocity plus the FWHM.  The single contour is drawn at
$N_{\rm HI}= 1.5\times10^{18}$cm$^{-2}$.~~{\it right:} Brightness
temperature spectra of CHVC\,186$+$19$-$114 at the indicated positions.
Solid lines refer to spectra from the WSRT, dashed lines to spectra
from the Arecibo telescope.  }\label{fig:h186a}
\end{figure*}

\begin{figure*}
\begin{tabular}{ccc}
\end{tabular}
\caption{CHVC\,186$+$19$-$114 as observed with the WSRT at a spatial
resolution of $2\times1$~arcmin and a kinematic resolution of
$2\rm\;km\;s^{-1}$.~~{\it left:} Apparent integrated HI in units of
$\rm K\;km\;s^{-1}$, with contours drawn
at~$10,\,20,\,\ldots,\,80\rm\;K\;km\;s^{-1}$, and with a linear
grey--scale extending from~$-1$ to~$68\rm\;K\;km\;s^{-1}$.~~{\it
middle:} Intensity--weighted line--of--sight velocity, $V_{\rm LSR}$,
with contours at $-120,\,-118,\,\ldots,\,-110\rm\;km\;s^{-1}$ and
linear grey--scale extending from~$-122$
to~$-109\rm\;km\;s^{-1}$.~~{\it right:} Distribution of the
intensity--weighted velocity dispersion, with a linear grey--scale
extending from~$0$ to~$6\rm\;km\;s^{-1}$. The contours correspond to
dispersions of 2 and 4~\kms, respectively.  }\label{fig:h186b}
\end{figure*}
 
\begin{figure*}
\begin{center}
\end{center}
\caption{Overlay of WSRT and Arecibo intensity data for
CHVC\,186$+$19$-$114. The WSRT intensities detected at a resolution of
$1\times2$~arcmin are indicated by contours drawn at
$1,\,3,\,5,\,7\rm\;K$.  The Arecibo data at 3 arcmin resolution are
indicated by the grey--scale images, with the scale given by the color
bar in units of Kelvin. }\label{fig:h186d}
\end{figure*}

\begin{figure*}
\begin{tabular}{cc}
\end{tabular}
\caption{~~{\it left:} Overlay of WSRT and HIPASS \NH~data for
CHVC\,148$+$82$-$258. The WSRT detected \hi column density at
$1.5\times8.5$~arcmin resolution is indicated by contours which are
drawn at $2,\,3,\,4\dots 9\times10^{18}$cm$^{-2}$.  The 15.5~arcmin
HIPASS data are indicated by the grey--scale images in units of
$10^{18}$cm$^{-2}$.~~{\it right:} Brightness temperature spectra of
CHVC\,148$-$82$-$258, measured with the WSRT at the indicated
positions.  }\label{fig:h148a}
\end{figure*} 

\begin{figure*}
\begin{tabular}{cc}
\end{tabular}
\caption{~~{\it left:} Overlay of WSRT and HIPASS \NH~data for
CHVC\,358$+$12$-$137. The WSRT detected \hi column density at
$1\times8$~arcmin resolution is indicated by contours which are drawn
at $2,\,3,\,4\dots 9\times10^{18}$cm$^{-2}$.  The 15.5~arcmin HIPASS
data are indicated by the grey--scale images in units of
$10^{18}$cm$^{-2}$.~~{\it right:} Brightness temperature spectra of
CHVC\,358$+$12$-$137, measured with the WSRT at the indicated
positions.  }\label{fig:h358a}
\end{figure*}


\begin{thebibliography}{}

 
\bibitem[2001]{barnes01}
Barnes D.\,G.,  Staveley--Smith L.,  de Blok W.\,J.\,G.,  et al., 2001,
MNRAS,
322, 486  
\bibitem[1999]{blitz99}
    Blitz~L., Spergel~D.\,N., Teuben~P.\,J., Hartmann~D., Burton~W.\,B.,
    1999, ApJ, 514, 818

\bibitem[2001]{burton01}
    Burton~W.\,B., Braun~R., Chengalur~J.\,N.,
    2001, A\&A, 369, 616

\bibitem[1992]{braun92}
    Braun~R., Walterbos~R.\,A.\,M.,
    1992, ApJ, 386, 120

\bibitem[1995]{braun95}
    Braun~R.,
    1995, A\&AS, 114, 409

\bibitem[1999]{braun97}
    Braun~R., 
    1997, ApJ, 484, 637

\bibitem[1999]{braun99}
    Braun~R., Burton~W.\,B.,
    1999, A\&A, 341, 437

\bibitem[2000]{braun00}
    Braun~R., Burton~W.\,B.,
    2000, A\&A, 354, 853

\bibitem[1980]{bregman80}
    Bregman J.\,N.,
    1980, ApJ, 236, 577

\bibitem[2001]{bruens01}
    Br\"uns~C., Kerp~J., Pagels~A.,
    2001, A\&A, 370, 26

\bibitem[1993]{corbelli93}
Corbelli E., Salpeter E.\,E., 1993, ApJ, 419, 104

\bibitem[1975]{davies75}
    Davies~R.\,D.,
    1975, MNRAS, 170, 45

\bibitem[2000]{avillez00}
    de Avillez M.\,A., 2000, AP\&SS, 272, 23

\bibitem[2002a]{deheij02a}
    de\,Heij~V., Braun~R., Burton~W.\,B.,
    2002a, A\&A, submitted

\bibitem[2002b]{deheij02b}
    de\,Heij~V., Braun~R., Burton~W.\,B.,
    2002b, A\&A, submitted
\bibitem[1978]{draine} Draine  B.\,T., 1978, ApJS, 36, 595
\bibitem[1976]{eichler}
Eichler  D., 1976, ApJ, 208, 694
\bibitem[1976]{einasto}
Einasto J., Haud U., J\^oeveer M., \& Kaasik A., 1976, MNRAS, 177, 357

\bibitem[1969]{field} Field  G.\,B., Goldsmith  D.\,W., Habing  H.\,J.,
1969, ApJ, 155, L149
\bibitem[1973]{giovanelli}
Giovanelli R., Verschuur G.\,L., Cram, T., 1973, A\&AS, 12, 209
\bibitem[1997]{hartmann97}
    Hartmann~D., Burton~W.\,B.,
    1997, Atlas of Galactic Neutral Hydrogen, Cambridge University Press
\bibitem[1978]{hulsbosch}
Hulsbosch A.\,N.\,M., 1978, A\&A, 66, L5
\bibitem[2001]{ibata01}
    Ibata R., Irwin M., Lewis G., Ferguson A.\,M.\,N., Tanvir N.,
    2001, Nature, 412, 49

\bibitem[1999]{klypin99}
    Klypin~A., Kravtsov~A.\,V., Valenzuela~O., Prada~F.,
    1999, ApJ, 522, 82

\bibitem[1993]{maloney93}
Maloney P., 1993, ApJ, 414, 41

\bibitem[1998]{mateo98}
    Mateo M., 
    1998, ARA\&A, 36, 435
\bibitem[1999]{moore99}
Moore  B., Ghigna  S., Governato  G., Lake  G., Quinn  T., Stadel  J.,
\& Tozzi  P. 1999, ApJ, 524, L19
\bibitem[1963]{muller63}
    Muller~C.\,A., Oort~J.\,H., Raimond~E.,
    1963, C.~R.~Acad.~Sci. Paris, 257, 1661

\bibitem[1978]{pellet} Pellet A., Astier N., Viale G., et al. 1978,
A\&AS, 31, 439

\bibitem[1999]{putman99}
    Putman~M.\,E., Gibson~B.\,K.,
    1999, PASA, 16, 70

\bibitem[2002]{putman02}
    Putman  M.\,E., de\,Heij  V., Stavely--Smith  L., et al. 2002, AJ,
123, 873
\bibitem[1976]{shap76}
Shapiro  P.\,R., \& Field  G.\,B. 1976, ApJ, 205, 762
\bibitem[1969]{verschuur}
Verschuur G.\,L., 1969, ApJ, 156, 771
\bibitem[2001]{wakker01}
    Wakker B.\,P.,
    2001, ApJS, 136, 537

\bibitem[1991]{wakker91b}
    Wakker~B.\,P., Schwarz~U.,
    1991, A\&A, 250, 484
\bibitem[1991]{wakker91}
    Wakker~B.\,P., van Woerden~H.,
    1991, A\&A, 250, 509

\bibitem[1997]{wakker97}
    Wakker~B.\,P., van Woerden~H.,
    1997, ARA\&A, 35, 217

\bibitem[1999]{wakker99}
    Wakker~B.\,P., van Woerden~H., Gibson~B.\,K.,
    1999, in: ASP Conf. Ser.~166, Stromlo Workshop on High--Velocity
    Clouds, B.\,K.~Gibson \& M.\,E.~Putman (eds.), p.~311
\bibitem[1991]{wakker91c}
    Wakker~B.\,P., Vijfschaft  B., Schwarz  U.\,J.
    1991, A\&A, 249, 233

\bibitem[1973]{williams}
Williams D.\,R.\,W., 1973, A\&AS, 8, 505
\bibitem[1999]{woerden99}
    van Woerden~H., Schwarz~U.\,J., Peletier~R.\,F., Wakker~B.\,P.,
    Kalberla~P.\,M.\,W., 1999, Nature, 400, 138
\bibitem[1995]{wolfire95}
        Wolfire M.\,G., McKee C.\,F., Hollenbach D., Tielens A.\,G.\,G.\,M.,
        1995, ApJ, 453, 673

\end{thebibliography}
\end{document}